\documentstyle[12pt]{article}

\begin{document}
\bibliographystyle{unsrt}

\begin{flushright} UMD-PP-95-31

August,1994
\end{flushright}

\vspace{6mm}

\begin{center}

{\Large \bf Implications of a Purely Right-handed $b$-decay Coupling}\\ [6mm]

\vspace{6mm}
{\bf{R.N. Mohapatra\footnote{Work supported by the
 National Science Foundation Grant PHY-9119745}~and~
S. Nussinov\footnote{Permanent address: Department of Physics and
Astronomy, Tel Aviv University, Tel Aviv, Israel.}}}

{\it{ Department of Physics and Astronomy,}}
{\it{University of Maryland,}}
{\it{ College Park, MD 20742 }}

\end{center}
\vspace{20mm}
\begin{center}
{\bf Abstract}
\end{center}
\vspace{6mm}

  We examine the implications of the hypothesis that the decays of the
bottom quark occur via pure right-handed couplings .
 We show that the existing lower limits on the lifetimes for neutrinoless
double beta decay severely restrict the neutrino sector of the model and would
 be satisfied in a natural manner if there is either an $L_e-L_\mu$ or
an $L_e-L_\tau$ symmetry in the model.
 We then show that the cosmological mass density constraints
 in combination with SN1987A observations imply lower bounds on
  the masses of the right-handed neutrinos of about
100 MeV ( about 65 MeV if there are Goldstone bosons coupling to $\nu_R$'s)
and also imply that the case of $L_e-L_\mu$ symmetry is inconsistent
with data unless there are flavor-changing neutral currents
 involving the right-handed neutrinos.

\newpage

It has recently been suggested by Gronau and Wakaizumi\cite{gronau}
 that unlike the quarks of the
first two generations, the decays of the bottom ( $b$ ) quark occur purely
via its right-handed couplings. In such models, the smallness of the
$b$ to $c$ coupling is attributed not to a small value of the corresponding
entry in the CKM matrix, $V_{cb}$ but rather to the smaller effective
right-handed Fermi coupling constant, $G^R_F$ compared
 to the usual left-handed Fermi coupling constant, $G^L_F$. Implementation
of this hypothesis
within the framework of the left-right symmetric models of weak interactions
\cite{moh} requires (i) that the left-handed quark mixing matrix be completely
different from the right-handed one; and (ii) that the right-handed W-bosons,
$W_R$ be rather light such that
$$G^R_F/G^L_F\simeq {{1}\over{\sqrt{2}}}
\left({{g^2_R}\over{M^2_{W_R}}}\right)/\left({{g^2_L}\over{M^2_{W_L}}}\right)
\equiv \left({{\beta_g}\over{\sqrt{2}}}\right)\simeq V_{bc}\simeq 0.04
\eqno(1)$$

In eq.(1), the factor ${{1}\over{\sqrt{2}}}$ is the maximal mixing angle
between the second and thirs generation in the right-handed charged currents.
Despite the overwhelming overall success of the standard $SU(2)_L\times U(1)_Y$
model, it has been shown that all B-decays, rare processes, $B-\bar{B}$ mixing
and CP-violation pattern can be reproduced in this model without excessive
fine tuning\cite{gronau} of parameters.
 Future special purpose experiments measuring
$\Lambda_b$ polarization via its weak decays\cite{rosner} will directly
settle this issue.

In this note, we would like to point out that considerations of the neutrino
sector , briefly alluded to already in the original paper\cite{gronau},
 very strongly constrain this model. As noted there, the low value for the
$M_{W_R}$ requires that neutrinos must be Majorana fermions. In
particular to be consistent with the results of experiments searching
for right-handed currents in muon and beta decay\cite{herczeg}
the right-handed neutrinos of the electron and muon generation must
be heavier than 7 MeV or so. On the other hand, the fact that in this
model, all semileptonic decays of the $b$-quark must proceed dominantly via
the emission of right-handed neutrinos implies an experimental
upper limit on the $m_{\nu_{iR}}$ ($i=e,\mu$)\cite{fulton}
\footnote{ In ref.\cite{gronau}, a stricter upper limit of about 200 MeV
has been used.}.
$$m_{\nu_{iR}}\leq 1~GeV\eqno(2)$$

Note that this is very different from the conventional left-right models
with the see-saw mechanism where $m_{\nu_R}\approx M_{W_R}$\cite{mohsen}
and as a result $\nu_R$ essentially decouples from the low energy effective
theory. In the model of ref.\cite{gronau}, the $\nu_R$'s are very much
a part of low energy world and will effect the low energy observations.

The first low energy process we analyze is the neutrinoless double beta
decay .
We  show that present limits on the lifetimes for neutrinoless
double beta decay\cite{dbb} can be satisfied in the model only if there is
extreme fine tuning in the $\nu_R$ sector of the theory or if there
is an exact $L_e-L_\mu$ or $L_e-L_\tau$ symmetry in the theory.

We then discuss the constraints from cosmology on this model.
Since the right-handed neutrinos in this model are in the moderate
mass range of less than a GeV, their relic density can violate
the bounds from  cosmological energy density\cite{kolb}
 as well as the structure formation\cite{steig} unless they decay
fast enough. Therefore
any model apart from satisfying the bounds from $\beta\beta 0\nu$
lifetimes, must also provide a mechanism for fast decay of the right-handed
neutrinos . Using these constraints in combination with the ones arising from
SN1987A observations on the decay mode $\nu_R\rightarrow e^+e^-\nu_L$,
 we  show that there must be a lower bound
 of  100 MeV on the mass
of the right-handed neutrinos, if they do not couple to any Goldstone bosons
and 66 MeV if they do.
In particular, we show that
that the existing ALEPH data\cite{aleph} which observed the
decay of the $b$-hadrons to  tau leptons  rules out  the first
possibility ( i.e. $L_e-L_{\mu}$ symmetry) in the context of
left-right models ( without Goldstone bosons )
provided there are no  flavor-changing neutral currents
involving the right-handed neutrinos.

\vspace{6mm}
\noindent{\it I. Minimal left-right model for pure right-handed $b$-decay:}
\vspace{6mm}

The minimal left-right gauge model that
leads to a pure right-handed $b$-decay couplings is a
slightly extended version of the   model discussed
in ref.\cite{mohsen} .
 The neutrinos in the model of ref.\cite{mohsen}
 are Majorana particles and the mass matrix involving
 the left and the right-handed neutrinos have the see-saw form
which for natural values of parameters leads to $m_{\nu_R}\approx M_{W_R}$
so that $\nu_R$ decouples from the low energy physics. Since in the model
under discussion, the $\nu_R$'s play a key role in the $b$-decays,
the parameters of the see-saw matrix have to be fine-tuned such that the
right-handed neutrinos lighter than 1 GeV. Secondly, the model of
ref.\cite{mohsen} due to the constraint of parity symmetry can lead
to symmetric quark and lepton mass matrices ( in the case that
 CP-violation is spontaneous) and this leads to identical quark
mixings in the left and right-handed charged currents.
We have to make sure that is not obeyed by the pure right-handed
$b$-decay model.

We denote the
 lepton fields by $\Psi_a \equiv {\pmatrix{\nu \cr e \cr }}_a$,
and the quark fields  by $Q_a\equiv {\pmatrix{u \cr d \cr}}_a$,
where $a~ =~ 1,~ 2,~ 3$.  Under the
gauge group $SU(2)_L \times SU(2)_R \times U(1)_{B-L}$, they transform as
$\Psi_{a~L} \equiv (1/2, ~ 0 , ~ -1 )$
and $\Psi_{a~R} \equiv (0, ~ 1/2, ~ -1 )$ and similarly for the quarks
with the appropriate $B-L$ quantum number.
 The Higgs sector
of the model consists
of the bi-doublet field
$\phi \equiv (1/2, ~ 1/2, ~ 0)$ and triplet Higgs fields:
${\Delta_L ( 1, ~0, ~ +2 ) \oplus \Delta_R (0, ~ 1, ~ +2 )
{}~~~~~}$. The Lagrangian for this model is given in ref.\cite{mohsen}.
The gauge symmetry is spontaneously broken by the vacuum expectation
values:${< {\Delta_R^0} > = V_R ~~; }$ ,
${< \Delta_L^0 > =  0 ~~;}$
 and $diag.{< \phi >} = ( { \kappa, \kappa^\prime }) $.
 As usual, $< \phi >$ gives masses to the charged fermions and Dirac masses
to the neutrinos whereas
$< \Delta_R^0 >$ breaks the $SU(2)_R$ symmetry and gives Majorana mass
to the right-handed neutrinos.

The charged current weak interaction Lagrangian in this model
has the following general form:
$$L_{wk}~=~{{g_L}\over{\sqrt{2}}}\left(\bar{P}_L \gamma_{\mu}V_LN_L
{}~+~\bar{\nu}_L \gamma_{\mu}U_L E_L\right)W^{\mu}_L~ $$
$$~~~~~~~+~{{g_R}\over{{\sqrt{2}}}}\left(\bar{P}_R \gamma_{\mu}V_RN_R
{}~+~\bar{\nu}_R \gamma_{\mu} U_R E_R\right)W^{\mu}_R~$$
$$+ \zeta_g M^2_{W_R}W^+_L W^-_R~+~h.c.\eqno(3)$$

In eq.(3), $P\equiv (u,c,t);~ N\equiv (d,s,b);~ \nu\equiv (\nu_e,\nu_{\mu},
\nu_{\tau})~and~E\equiv (e,\mu,\tau)$. The $V_{L,R}$ and $U_{L,R}$ denote
the weak mixing matrices in the quark and lepton sectors respectively.
In order to have $V_L\neq V_R$, we must have either hard CP-violation
with the vev's of the bidoublet fields complex or we can have
spontaneous CP-violation with two bi-doublet fields\cite{mohc}.
The gauge mixing parameter $\zeta_g$ is bounded by $K\rightarrow 3\pi$
decay to be less than $0.01$.

The right-handed $b$-decay model is characterised by the following
 $V_L$ and $V_R$\cite{gronau}:
$$V_L~=\left(\begin{array}{ccc}
cos\theta_c & sin\theta_c & 0 \\
-sin\theta_c & cos\theta_c & 0 \\
0 & 0 & 1 \end{array}\right) ;~~
V_R~\simeq \left(\begin{array}{ccc}
1 & -s & s \\
{{s}^3\over{2\sqrt{2}}} & {{1}\over{\sqrt{2}}} & {{1}\over{\sqrt{2}}}\\
-s\sqrt{2} & -{{1}\over{\sqrt{2}}} &  {{1}\over{\sqrt{2}}} \end{array}
\right) \eqno(4)$$

We have ignored the CP-violating phase in $V_R$. Eq.(4) implies
that the $b$-quark decays via the right-handed currents.
Here, $\theta_c$ is the Cabibbo angle and
the parameter $s= |V_{ub}/V_{cb}|\sqrt{2}\simeq 0.08\pm 0.03$.
 After the fermion mass matrices
are diagonalized, the model will lead to the weak interaction Lagrangian
of eq.(3). There are no Goldstone bosons in the model.
 We will also assume that all left-handed
neutrinos are Majorana particles and
 have masses in the electron-volt range so that they are consistent
with the laboratory upper limits as well as the
cosmological constraints.

\vspace{6mm}
\noindent{\it II. Bounds from neutrinoless double beta decay:}
\vspace{6mm}

Let us now consider the implications of the present limits on the
lifetimes for neutrinoless double beta decay\cite{dbb} for this
 model.There are several contributions to $\beta\beta 0\nu$ in this
model: (i) left-handed $\nu$ exchange; (ii) right-handed $\nu_R$ exchange;
(iii) $\nu_L-\nu_R$ mixing and (iv) $\Delta^{++}_R$ exchange.
The contributions (i) and (iv) can be adjusted to obey the experimental
limits without effecting the main idea of the model. Therefore , we
focus on the  right-handed neutrinos and the $\nu_L-\nu_R$  mixings.
As already noted in ref.\cite{gronau}, the $\nu_{{e,\mu}R}$ must have masses
above 7 MeV  in order to be consistent with weak decay data.
Our main observation in this paper is that since they are Majorana
fermions and since the $W_R$ mass is low, their contribution to neutrinoless
double beta decay is non-negligible. In fact, if we write the rate for
$\beta\beta 0\nu$ decay process for a nucleus such as $^{76}Ge$ as:
$$\Gamma(\beta\beta 0\nu)\simeq {{G^4_F Q^5 }\over{60\pi^3}}|A|^2\eqno(5)$$
 $A$ will be given by\cite{burgess}:
$$|A|\simeq {{\omega_0 p_F E_F\beta^2_g}\over{4\pi^3}}\Sigma_i U^2_{Rei}
m_{\nu_{iR}}
{}~~~~~~~~~(for~~m_{\nu_{iR}}\ll~E_F\ll~p_F~)\eqno(6a)$$
and
$$|A|\simeq {{\omega_0 E_F p^3_F\beta^2_g}\over{12\pi^3}}\Sigma_i{{U^2_{Rei}}
\over{m_{\nu_{iR}}}}~~~~~~~~~~( for~~E_F\ll p_F\ll m_{\nu_{iR}})\eqno(6b)$$

In eq.(6), $p_F$ and $E_F$ denote the Fermi Momentum and Fermi energy of the
nucleons in the decaying nucleus ; $p_F\approx 100 MeV$ and $E_F\approx 5~MeV$
for typical nuclei of interest. Therefore, the first equation eq.(6a)
applies when the relevant $\nu_R$  lighter than 100 MeV and the
eq.(6b) applies when in the opposite case. In ref.\cite{burgess}, the
parameter $\omega_0$ was estimated from two neutrino double beta decay
to be about $4~MeV^{-1}$, which we use here.
 The present neutrinoless double beta decay experiments\cite{dbb}
imply the following bounds on the parameters of the model:

Case(a):
$$\beta^2_g\Sigma_i U^2_{Rei} m_{\nu_{iR}}\leq 2~eV~~~for~7~MeV\leq~m_{\nu_R}
\leq 10~MeV~\eqno(7a)$$

Case(b):
$$\beta^2_g\Sigma_i{{U^2_{Rei} p^2_F}\over{3 m_{\nu_{iR}}}}\leq 2~eV~~~~
for~.1~GeV\leq m_{\nu_R}\leq~1~GeV\eqno(7b)$$

According to eq.(1), the parameter $\beta_g\simeq .055$ and
 the inequalities (7a) and (7b) are violated by four orders of magnitude,
if we do not assume any fine tuning among the parameters
in the sum. In the intermediate region of $10~MeV\leq m_{\nu_R}\leq 100~MeV$,
we expect a similar conclusion to hold.( We have used
 the fact that the semi-leptonic
decays of the $b$-quark require the mixing angle $U_{Ree}\simeq 1$).
This leads us to conclude that {\it the only way to make the right-handed
$b$-decay model consistent with limits from $\beta\beta 0\nu$ is to
demand severe fine tuning in the right-handed neutrino sector so that
in case(a), $\Sigma_i U^2_{Rei} m_{\nu_{iR}}\leq keV$ and a similar strong
cancellation for the heavy $\nu_R$ case (b)}.

One way to achieve this cancellation is to have an exact $L_e-L_\mu$
or $L_e-L_\tau$ symmetry in the theory so that the neutrino less double beta
decay is forbidden by the symmetry.

\vspace{6mm}
\noindent{\it III. Cosmological constraints:}
\vspace{6mm}

Let us now turn to the discussion of the cosmological constraints
on the model.
In order to see how the heavy right-handed neutrinos can avoid the
cosmological mass density constraints, let us first consider the
effects of annihilation in the early universe. Since the $W_R$ and $Z^{\prime}$
are only a few times heavier than the usual $W$'s and $Z$'s, the discussion
 of Lee and Weinberg\cite{lee} for relic abundance of heavy neutrinos
can be carried over with small modifications and one must require $m_{\nu_R}
\geq 6~GeV$ in order to satisfy the mass density constraints. Since
semileptonic $b$-decays to all three leptons have been observed,
the right-handed neutrinos cannot satisfy this bound.
We must therefore find a mechanism for the right-handed neutrinos to decay
fast enough in the model. Since the $\nu_R$'s decouple in the
early universe when they are non-relativistic, their masses and lifetimes
must satisfy the constraint\cite{kolb}:
$${{1}\over{\beta^{{\prime}{2}}_g}}\left({{1~GeV}\over{M_{\nu_{iR}}}}\right)^2
\left({{\tau_{\nu_R}}\over{t_U}}\right)^{{1}\over{2}}\leq .05 h^2\eqno(8)$$

Here, $h$ is the Hubble constant in units of $100~km~sec.^{-1}(Mpc)^{-1}$.
In eq.(8), $\beta^{\prime}_g\equiv \left(g^2_R M^2_Z/ g^2_L M^2_{Z^{\prime}}
\right)$ since the $\nu_R$ annihilation takes place via the exchange of
the heavier $Z^{\prime}$ in the model. In the minimal left-right model
described above, we have the relation that $M^2_{Z^{\prime}}\simeq
3 M^2_{W_R}$\cite{mohsen} leading to
$\beta^{\prime}_g\approx .42\beta_g$ .

\vspace{4mm}
\noindent{\it $L_e-L_\mu$ symmetry:}
\vspace{4mm}

 To see the implications of eq.(8) for the $\nu_R$ spectrum, let us
first consider the model with an $L_e-L_\mu$ symmetry.
Due to the $L_e-L_\mu$ symmetry of the model, the $\nu_{\mu_R}$
and the $\nu_{eR}$ are degenerate and and their decay can occur
either via pure $W_R$ exchange or via $W_R-W_L$ mixing leading
to the following final states:

\noindent{\underline{\it  $W_L-W_R$ mixing induced decays:}}

{}~~~(a):~$\nu_{e,\mu R}\rightarrow e^+e^-\nu_e$~~~~~for $m_{\nu_R}\leq
108~MeV$

{}~~~(b):~$\nu_{e,\mu R}\rightarrow e^{-}\mu^{+}\nu_{\mu}$~~~~~ for
$108~MeV\leq m_{\nu_R}\leq 140~ MeV$;

{}~~~(c):~$\nu_{e,\mu R}\rightarrow e^{\pm}\pi,~~ e^\pm~hadrons$~~~~~ for
 $m_{\nu_R}\geq 140$ MeV.

\noindent{\underline{\it $W_R$ induced decays:}}

In this case, if $m_{\nu_R}$ is less than 140 MeV or so, there are no
kinematically allowed decay modes; as a result, the $\nu_R$ stable. But,
it is more massive than 140 MeV, the hadronic decay modes open up so that
$\nu_R$ becomes unstable.

{}From eq.(8) and the the decay rate $\tau^{-1}_{\nu_R}\simeq
G^2_F\zeta^2_g m^5_{\nu_R}/(192\pi^3)$ we find
 that $m_{\nu_{e R}}\geq 30$ MeV for
 $\zeta_g\leq .01$. However, we will show in a subsequent section
 that there are very strong bounds on the $\nu_R\rightarrow e^+e^-\nu_e$ mode
from lack of observation of gamma rays by the SMM  during
SN1987A \cite{mnz} , which help to push the lower bound to  100 MeV.
  For the choice of parameters of the model, the other cases
all satisfy the constraint in eq.(8) so that the lower bounds respectively
are 110 MeV for the case of $\zeta_g\neq 0$ and 140-150 MeV otherwise.

Coming now to the $\nu_{\tau R}$  decay,
it can occur via three different mechanisms;(i) the flavor changing
neutral current (FCNC) interactions induced via the see-saw form for the
neutrino masses; (ii)  the $W_L-W_R$ mixing and (iii) via $W_R$ exchange.
In the FCNC mechanism, the typical
strength of the interaction is $\approx G_F\times\zeta_{\nu}$;
for $m_{\nu_{\tau_R}}\leq 140$ MeV,
the final state is $e^+e^{-}\nu_{\tau_L}$; otherwise, the single pion and
other final states open up.
Using eq.(8), and an analogous decay rate formula, we  get:
$$\left({{m_{\nu_{\tau_R}}}\over{1~GeV}}\right)^9\zeta^2_\nu\geq .72\times
10^{-18}\eqno(9)$$
One can obtain a plausible estimate for
 $\zeta_\nu$ from the see-saw formula for neutrinos to be:
  $\zeta^2_{\nu}\approx {{m_{\nu_{\tau L}}}/
{m_{\nu_{\tau R}}}}\approx  10^{-3}$ .
Eq.(9) then
 implies that for this mechanism to produce a decay fast enough for
cosmology, we must have $m_{\nu_\tau R}\geq 45~MeV $ or so.
But again , since the only final state here is the $e^+e^-\nu_\tau$,
SN1987A bounds will apply and the lower limit on $m_{\nu_{\tau_R}}$
will become 100 MeV or so.

On the other hand, if $W_L-W_R$ mixing or $W_R$ exchange provide
 the dominant decay mechanism, then the only
allowed final state involves  a $\tau$ lepton and therefore, we must have
$m_{\nu_{\tau}}\geq 2$ GeV or so for this decay to be kinematically
allowed. This is however inconsistent with  data\cite{fulton} on
semileptonic decays ruling out $L_e-L_{\mu}$ symmetry as a way to satisfy
 the $\beta\beta 0\nu$ constraints.

\newpage
\noindent{\it $L_e-L_{\tau}$ symmetry:}
\vspace{4mm}

For the case with $L_e-L_\tau$ symmetry,
 we have ${\nu_{eR}}$ and ${\nu_{\tau R}}$ degenerate.
So, the  $\nu_{eR}$ ( or $\nu_{\tau R}$) can  decay as in the
previous case and the bounds cited above apply. As for the $\nu_\mu$,
it can decay either via the FCNC mechanism in an analogous manner
to the above or via the $W_L-W_R$ mixing or the $W_R$ exchange.
In first case, the most stringent lower bound will arise from SN1987A
gamma ray observations.
 In the other cases,
since a muon must be produced in the final state, the bound increases
to 110 MeV or so.

\vspace{6mm}
\noindent{\it IV. Bounds from supernova cooling:}
\vspace{6mm}

Finally, we briefly comment on the constraints on the model
from supernova cooling.
In the usual scenario\cite{burrows} for supernova cooling, the 1-10 sec.
cooling period of the hot collapsed core is via $\nu_{iL}$ emission.
Observation of the neutrino burst from the supernova SN1987A\cite{hirata}
confirm this idea. A simple minded way to understand this is to note that
the effective mean free path of a typical neutrino  ( with $E_\nu\simeq
30~MeV$) is given by $\lambda_\nu \simeq \left( n_T \sigma^{eff}_\nu\right)^
{-1}\approx \left(n_T G^2_F E^2_{\nu}\right)^{-1}$. If we use $n_T\approx
10^{39}$ corresponding to a core density $\rho_c\approx 10^{15} gr.(cm)^{-3}$,
then we get $\lambda_\nu\approx 30~ cm$. In the random walk approximation, the
total diffusive path length is given by $\ell_{tot}={{R}^2\over{\lambda_\nu}}
\approx 3\times 10^{8} meters$ for $R\approx 10~km$. Since
 the neutrinos travel with the velocity
of light, this implies that the neutrinos escape the supernova in about
a second ; this gives the cooling time $\Delta t\simeq 1~sec.$.

In the present scenario, the production of right-handed neutrinos
in the supernova core takes
place via the following mechanisms\cite{barb} : (i) $e_R+p\rightarrow n+\nu_R$;
(ii)$e^+_R+n\rightarrow p+\nu_R$ and (iii) $e^+e^-\rightarrow \nu_R \nu_R$.
These production rates as well as other interaction rates for the $\nu_R$'s
  are smaller than those for the left-handed neutrinos
due to higher mass for the $W_R$'s . For $M_{W_R}\approx 5\times M_{W_L}$,
the mean free path for the $\nu_R$'s is longer by a factor of roughly $\approx
600$ . This implies that the time required per collision is $\lambda_{\nu_R}
/c\approx 10^{-6} sec.$ As a rseult,
 the $\nu_R$'s formed travel out of the core
much faster than the $\nu_L$'s.
 Thus the energy loss via $\nu_R$ diffusion can be thought of
qualitatively as a two step process- (i) formation on a short distance
scale due to electron mean free path being small and (ii) diffusion of
the $\nu_R$ over a much further distance before a recapture- this distance
 being about 600 times the corresponding $\nu_L$ travel path. Repeating this
process, leads to the $\nu_R$ cooling mechanism with an effective time scale
 for this is of order:
$$\Delta t^{e}_R\approx \Delta t_L \left({{\lambda_{\nu}}
\over{\lambda_{\nu_R}}}
\right)\leq 10^{-3}sec.\eqno(10)$$
 For the $\nu_{\mu,\tau R}$
cases, there are only $Z^{\prime}$ contributions and therefore the
corresponding $\Delta t_R$ is of order $10^{-4}$ sec.
These qualitative considerations make it clear that
 $\nu_R$ cooling is much more efficient  contrary to observation.
Clearly if  $m_{\nu_R}\geq 35$ MeV or so, this additional cooling
mechanism will be Boltzman suppressed. Let us note
that, this bound is much less stringent than the ones derived earlier.

\vspace{6mm}
\noindent{\it V. Bounds on the $e^+e^-\nu_L$ decay mode from $\gamma$-ray
data from SN1987A:}
\vspace{6mm}

We saw in sec.III that there are several situations where the $\nu_R$ can
decay to $e^+e^-\nu_L$ final states and that for allowed ranges of parameters
in the theory, the cosmological mass density constraints allow a lower bound
in the range of 30 to 55 MeV. In this section, we will show that in these
cases, lack of any evidence for $\gamma$-rays in the MeV range from the
Solar Maximum Mission\cite{chupp} implies a more stringent lower bound.
First note that for supernova emission rate for
heavier right-handed neutrinos is suppressed by the Boltzman factor
$e^{-{{m_{\nu_R}}\over{kT}}}$. If the only ( or dominant) decay mode for
$\nu_R$ involves the $e^+e^-$ final state, then these final state products
can lead to gamma rays either via bremstrahlung or annihilation. The
details of this were discussed in ref.\cite{mnz}, where it was shown
that if a heavy neutrino lives longer than about 100 sec. and it decays
to either $e^+e^-\nu_L$ or $e^+e^-\nu_L\gamma$, then there will be
an observable photon flux near the earth.
 For us the relevant
gamma-ray production mechanism is the radiative decay$\nu_R\rightarrow
e^+e^-\nu_L \gamma$. The local photon flux due to this mechanism was estimated
in the first reference of \cite{mnz} to be:
$$\Phi_{\gamma}\simeq {{\alpha N_{\nu_R} B}\over{8\pi^2 D^2 \Delta
t}}\eqno(11)$$

In the above equation, $N_{\nu_R}$ stands for the number of $\nu_R$'s emitted
from the supernova; B denotes the branching ratio for the decay mode $\nu_R
\rightarrow e^+e^-\nu_L$ and $\Delta t$ denotes the duration of neutrino
emission in the supernova ( taken to be 10 sec. below); D is the distance
from SN1987A to the earth.
Experimentally, $\Phi_{\gamma}\leq 0.1~ cm^{-2} sec^{-1}.$.
We will assume that the
entire suppression in the gamma ray signal is due to
 the Boltzman suppression of the emission  intensity of $\nu_R$.
We have two sources for this suppression: (i) one is the Boltzman
suppression factor for the higher mass and (ii)the other arises from the
fact that for the parameters of our model, the typical lifetimes for
$\nu_R$ are about 5-10 sec. Combining them,
  we get (for $B=1$),
$$\left({{m_{\nu_R}}\over{kT}}\right)^{{3}\over{2}}
e^{-{{m_{\nu_R}}\over{kT}}}e^{-{{t}\over{\tau}}}
\leq {{10^{-10}}/({m_{\nu_R}~in~MeV})}\eqno(12)$$

Assuming the temperature of the emitted $\nu_R$'s to be higher
than 5  MeV, $t=100~sec.$, $\tau = 10~sec.$,
we get $m_{\nu_R}\geq~100$ MeV.

The lower bound derived in eq.(12) clearly deteriorates as $\tau_{\nu_R}$
becomes much shorter than 100 sec. However, it is possible to derive
a bound which is valid as long as $\tau_{\nu_R}\geq 10^{-3}~sec.$ (which
is the time required for  $\nu_R$ to escape the core ).

Let us assume that $\tau_{\nu_R}\leq 100~sec.$ so that the majority of
the $\nu_{iR}$'s decay inside the progenitor of SN1987A. This decay,
if it involves $e^+e^-$ final states will deposite inside the mantle
of magnitude $\approx W_{\nu R}$, the total energy carried by the $\nu_R$'s.
$W_{\nu R}$ is suppressed by the Boltzman factor but is enhanced by the
larger core escape efficiency of the $\nu_R$'s; so we estimate:
$$W_{\nu R}\approx ({{m_{\nu_R}}\over{T}})^{{3}\over{2}}
e^{-{{m_{\nu_R}}\over{T}}}
\times({{\Delta t_{\nu_L}}\over{\Delta t_{\nu_R}}})W_{\nu L}\eqno(13)$$
$W_{\nu L}$ is the total $\nu_L$ luminosity.
Demanding that the energy deposited will not exceed the total mantle
( explosion)energy $\approx 10^{51}~ergs\approx 10^{-2}W_{\nu L}$ implies that:
$$({{m_{\nu_R}}\over{T}})^{{3}\over{2}}e^{-{{m_{\nu_R}}\over{T}}}\leq
10^{-5}-10^{-6}\eqno(14)$$
 For $T_{\nu_R}\approx 5$ MeV, we get from eq.(14),
 $m_{\nu_R}\geq 75- 90$ MeV.

Note that the above considerations referring only to the total energy
deposition are very conservative.In the conventional mechanism the
above $10^{51}$ erg energy is transferred to the mantle via the shock wave
and takes several hours to reach the stellar surface resulting in the observed
several hour delay between the time of the neutrino pulse and the supernova
explosion. In the case of the $\nu_R$ decay, the time of energy depostion
is of order of 100 sec rather than a few hours. Although without a detailed
modelling of the supernova, it is hard to say precisely how much this will
improve the bound , it is none-the-less clear that it would imply the
right hand side of eq.(14) to be even less and will lead to stronger
bounds on $\nu_R$ masses.

\vspace{6mm}
\noindent{\it VI. Goldstone boson coupling to right-handed neutrinos:}
\vspace{6mm}

Next, let us  consider the possibility that the right-handed neutrinos
decay via their couplings to a Goldstone boson invoked in ref.\cite{gronau}
 for the purpose of allowing faster two-body decays in order to avoid
cosmological constraints.
 The detailed gauge model
in this case will have to be more elaborate than the one mentioned
( e.g. ref.\cite{alok} as one possibility).
It appears however that, one can still
draw similar lower bound on the right-handed neutrinos in this
case in a fairly model independent manner. The constraints arise from
the interplay between the cosmological constraint in eq.(8) and the
requirement that the coupling of any Goldstone boson to electrons must
have a strength less $\leq 10^{-12}$ in order to satisfy the
constraints from the cooling of red giants\cite{dicus} .
  Denoting the  $\nu_R\rightarrow \nu_L + \chi$
(where $\chi$ is the massless Goldstone boson) coupling by $f_{\nu_R \nu_L
\chi}$, we get from eq.(8) the following inequality:
$$m^5_{\nu_R}f^2_{\nu_R \nu_L\chi}\geq 1.2\times 10^{-30}~GeV^5\eqno(15)$$

 The value of $f_{\nu_R \nu_L\chi}$ is expected to be
 of the same order of magnitude as the
$e e\chi$ coupling since both of them arise from the
same Yukawa interaction in the basic
 Lagrangian i.e.the $\bar{\psi}_L\phi\psi_R$
term. Therefore, using the astrophysical arguments, we
can assume the $f_{\nu_R \nu_L \chi}$
to be also $\approx 10^{-12}$. Then , in order to satisfy
 eq.(8), we need $m_{\nu_R}\geq 66$ MeV .

In conclusion, we have pointed out that in the model for pure right-handed
$b$-quark decay model, constraints of neutrinoless double beta decay
and cosmology require that there must either be an $L_e-L_\mu$ or
$L_e-L_{\tau}$ symmetry in the model. Furthermore, if there are no flavor
changing neutral current interactions involving the right-handed
neutrino in the model, then only symmetry
that works is the $L_e-L_\tau$. In either of the cases , we find that
cosmology and SN1987A observations  imply
 lower bounds on the right-handed neutrino masses
which are significantly higher than the 7 MeV value used in ref.\cite{gronau}.
We summariza these bounds for various cases in table I.
 Future high precision
 measurements of the semileptonic
$B\rightarrow D^*\ell \nu_R$ decays sensitive to neutrino masses
in the range of 100 MeV will provide decisive tests of this model.
In any case, if we combine our lower bounds with the upper bound of
200 MeV used in ref.\cite{gronau}, the allowed window for the model
gets considerably narrowed.

\vspace{6mm}
\begin{center}
{\bf Acknowledgement}
\end{center}
\vspace{6mm}

We would like to thank A. Jawahery  for many useful
discussions on $b$-decays and one of us (S. N.)would like to thank
the Nuclear Theory group of the University of
Maryland for kind hospitality.

\begin{center}
{\bf Table I}
\end{center}
\begin{tabular}{|c||c||c||c||c||c|}  \hline
Particle &  Symmetry & Decay mechanism & final state & $(m_{\nu_R})_{min}$
 & Source \\ \hline
$\nu_{e,\mu R}$ & $L_e-L_\mu$ & $W_L-W_R$ mixing & $e^+e^-\nu_L$&100 MeV
& SN1987A \\ \hline
$\nu_{\tau R}$ & " & $\nu_L-\nu_R$ mixing & "& 100 MeV & SN1987A\\ \hline
$\nu_{\tau R}$ & " & $W_L-W_R$ mixing  & $\tau+X$ & 2 GeV & cosmology\\
               &   & or $W_R$ exchange &      &     &  \\  \hline
$\nu_{e,\tau R}$ & $L_e-L_\tau$ & $W_L-W_R$ mixing &$e^+e^-\nu_L$&
 100 MeV & SN1987A\\ \hline
$\nu_{e,\tau R}$ & "& $W_R$ exchange & $e\pi$ & 140 MeV & cosmology\\ \hline
$\nu_{\mu R}$ & "& $W_L-W_R$ mixing & $\mu + X$& 110 MeV & cosmology \\ \hline
$\nu_{\mu R}$ & "& $\nu_L-\nu_R$ mixing & $e^+e^-\nu_L$ &100 MeV
 & SN1987A\\ \hline
all $\nu_R$ & any & $\nu_R\rightarrow \nu_L +\chi$ & $\nu_L +\chi$& 66 MeV
&
cosmology\\ \hline
\end{tabular}

\vspace{6mm}
\noindent{\bf Table caption:} The table summarizes the lower bounds on the
different $\nu_R$'s for various cases.

\end{document}